\documentclass[a4paper]{jpconf}

\usepackage{hyperref} 
\usepackage{amsmath}
\usepackage{amssymb}
\usepackage{graphicx}

\renewcommand{\d}{{\,\rm d}}

\newcommand{\meant}[1]{\left<#1\right>}
\newcommand{\meanx}[1]{\overline{#1}}

\DeclareGraphicsExtensions{.png,.pdf}
\graphicspath{{pictures/}}

\begin{document}

\title{Elements of sub-quantum thermodynamics: quantum motion as ballistic diffusion} 

\author{G Gr\"ossing, S Fussy, J Mesa Pascasio and H Schwabl}

\address{%
Austrian Institute for Nonlinear Studies\\
Akademiehof\\
Friedrichstr.~10, 1010 Vienna, Austria
}

\ead{ains@chello.at}

\begin{abstract}
By modelling quantum systems as emerging from a (classical) sub-quantum thermodynamics, the quantum mechanical ``decay of the wave packet'' is shown to simply result from sub-quantum diffusion with a specific diffusion coefficient varying in time due to a particle's changing thermal environment. It is thereby proven that free quantum motion strictly equals ballistic diffusion. The exact quantum mechanical trajectory distributions and the velocity field of the Gaussian wave packet are thus derived solely from classical physics. Moreover, also quantum motion in a linear (e.g., gravitational) potential is shown to equal said ballistic diffusion. Quantitative statements on the trajectories' characteristic behaviours are obtained which provide a detailed ``micro-causal'' explanation in full accordance with momentum conservation.
\end{abstract}

\section{Quantum mechanical dispersion of a free Gaussian wave packet exactly modelled by sub-quantum ballistic diffusion}
\label{sec:quantum}

Considering quantum theory as \textit{emergent} \cite{carroll.2010}, we propose that it may result from a deeper, more exact theory on a sub-quantum level. In our approach, one assumes that the latter can be described with the aid of non-equilibrium thermodynamics. We ask ourselves how quantum theory would have evolved, had the ``tool'' of modern non-equilibrium thermodynamics existed, say, a century ago. As one of us (G.\ G.) has recently shown, one can derive the exact Schr\"odinger equation with said tool, where the relation between energy $E$ and frequency $\omega $ is used as the only empirical input, $E= \hbar \omega $ \cite{Groessing.2008vacuum,Groessing.2009origin}, with the additional option that even the appearance of Planck's constant $\hbar $ may have its origin in classical physics \cite{Groessing.2010explan}. For an extensive review of refs.\ \cite{Groessing.2008vacuum} and \cite{Groessing.2009origin}, for connections to similar work, and, in particular, to Fisher information techniques, see \cite{carroll.2010}. As to approaches in a similar spirit, see, for example, \cite{adler.2004,nelson.1985,el_naschie.1995,nottale.1993,fritsche.2003,Elze.2009attractor,rusov.2009,tHooft.2001}.

In the thermodynamic approach to quantum behaviour \cite{Groessing.2008vacuum,Groessing.2009origin,Groessing.2010explan}, a particle of energy $E= \hbar \omega $ is characterized by an oscillator of angular frequency $\omega $, which itself is a dissipative system maintained in a non-equilibrium steady-state by a permanent troughput of energy, or heat flow, respectively. The latter is a form of kinetic energy different from the ``ordinary'' kinetic energy of the particle, as it represents an additional, stochastic contribution to it, i.e., from the presence of zero point fluctuations. Therefore, one deals with a context-dependent total energy of the whole system (i.e., the particle as the ``system of interest'' in a narrower sense and the heat flow constituting the particle's noisy thermal embedding), which is assumed as
\begin{equation} \label{eq:1.1} 
  E_\text{tot} = \hbar \omega + \frac{(\delta p)^2}{2m}\;, 
\end{equation} 
where $\delta p:=mu$ is said additional, environment-related fluctuating momentum of the particle of mass $m$.

We let ourselves be inspired by the experiments with the classical ``walkers'' of 
Yves Couder's group (see, for example, \cite{Protiere.2006,Couder.2006,Eddi.2009,Couder.2005}). One has to do with a rapidly ``bouncing'' (oscillating) object, which itself is guided by an environment that contributes some fluctuating momentum to the walker's propagation: a bouncing droplet phase-locked with an oscillating fluid surface. 
However, the walker is also the cause of waves emanating from the particle ``hitting'' the surface, and it is the detailed structure of the emerging wave configurations, in turn, that stochastically influences the walker's path. Thus, if we imagine the bouncing of a walker in its fluid environment, the latter will become ``excited'' or ``heated up'' wherever the momentum fluctuations direct the particle to. 
After some time span, a whole area of the particle's environment will be coherently ``heated up'' in this way. 
We shall here try to implement a similar ``particle-wave coupling'' with respect to a quantum system.

To start with a first major aspect of our modeling, we consider a particle of initial velocity $v$ in the midst of ``well ordered'' diffusion waves, which themselves emerge out of the noisy, Brownian-type diffusions of myriads of single sub-quantum particles. Being swept along with a diffusion wave, with initial ($t=0$) location $x(0)$ and diffusion velocity $u$, a thus described ``quantum'' particle's distance to some heat accumulation's center $x_0 $ at time $t$ will be
\begin{equation} \label{eq:1.12} 
  x(t) = x(0) + ut\;, 
\end{equation} 
such that one obtains the r.m.s.\ of \eqref{eq:1.12} as
\begin{equation} \label{eq:1.13} 
  \meanx{x^2(t)} = \meanx{x^2(0)} + 2 \; \meanx{x(0)\cdot u(x,t)} \; t + \meanx{u^2 (x,t)} \; t^2 \;. 
\end{equation}

Now we introduce an essential point: we assume, as an emerging result out of the statistics of a vast number of diffusion processes, the complete statistical independence 
\textit{at any time $t$}  of the velocities $u$ and the positions $x$, and thus also of $u$ and $v$: 
\begin{equation} \label{eq:1.14} 
  \meanx{xu} = 0 \quad \text{and} \quad \meanx{vu} = 0 \;. 
\end{equation} 
This is justified considering the statistics of huge numbers, millions of millions of diffusive sub-quantum Brownian motions, which are supposed to bring forth the emergence of said larger-scale collective phenomenon, i.e., the diffusion wave fields as solutions to the heat equation \cite{Groessing.2009origin}.  Therefore, with the \textit{average} orthogonality of classical (convective) momentum $mv$ on one hand, and its associated diffusive momentum $mu$ on the other, one gets rid of the term linear in $t$ in Eq.\eqref{eq:1.13}, and thus of irreversibility, and one obtains
\begin{equation} \label{eq:1.15} 
  \meanx{x^2(t)} = \meanx{x^2(0)} + \meanx{u^2} \, t^2 \;. 
\end{equation} 
Eq.~\eqref{eq:1.15} is the result obtained for the ``pure'' emergent diffusive motion as given by \eqref{eq:1.12}.

Now, let us assume we have a source of identical particles, which are prepared in such a way that each one ideally has an initial (classical) velocity $v$. Even if we let them emerge one at a time only, say, from an aperture with unsharp edges (thus avoiding diffraction effects to good approximation), the probability density $P$ will be a Gaussian one. This comes along with a heat distribution generated by the oscillating (``bouncing'') particle(s), with a maximum at the center of the aperture $x_0 =vt$. As the classical diffusion equation, in one dimension for simplicity, is given by
\begin{equation} \label{eq:1.2} 
  \frac{\partial}{\partial t} P(x,t) = D\frac{\partial^2}{\partial x^2} P(x,t)\;, 
\end{equation} 
where $D$ is the ``diffusion constant'', we have the corresponding solution
\begin{equation} \label{eq:1.3} 
  P(x,t) = \frac{1}{\sqrt{2\pi} \sigma} \e^{-\frac{(x-x_0 )^2 }{2\sigma^2 } }
\end{equation} 
with the usual variance $\sigma^2 = \meanx{x^2(t)}$. Note that from Eq.~\eqref{eq:1.1} one has for the \textit{averages} over particle positions and fluctuations (as represented via the probability density $P$)
\begin{equation} \label{eq:1.4} 
  \meanx{E_{{\rm tot}} } = \meanx{\hbar \omega }+\frac{\meanx{(\delta p)^2 }}{2m} = {\rm \; }{\rm const.}, 
\end{equation} 
with the mean values (generally defined in $n-$dimensional configuration space)
\begin{equation} \label{eq:1.5} 
  \meanx{(\delta p)^2 } := \int P (\delta p)^2 \d^n x \;. 
\end{equation} 
Equation \eqref{eq:1.4} is a statement of total average energy conservation. This means that a variation in $\delta p$ implies a varying ``particle energy'' $\hbar \omega $, and vice versa, such that each of the summands on the right hand side for itself is not conserved. In fact, as shall be detailed below, there will generally be an exchange of momentum between the two terms providing a net balance 
\begin{equation} \label{eq:1.6} 
  m\delta v = m\delta u \;, 
\end{equation} 
where $\delta v$ describes a change in the ``convective'' velocity $v$ paralleled by the ``diffusive'' momentum fluctuation $\delta (\delta p):=m\delta u$ in the thermal environment.

As elaborated in references \cite{Groessing.2008vacuum} and \cite{Groessing.2009origin}, once Eq.~\eqref{eq:1.1} is assumed, considerations based on Boltzmann's relation between action and angular frequency of an oscillator provide, without any further reference to quantum theory, that
\begin{equation} \label{eq:1.7} 
  \delta p = mu := \hbar k_{\rm u} = -\frac{\hbar }{2} \nabla \ln P\;.
\end{equation} 
Further, as can easily be checked by integration, it holds that
\begin{equation} \label{eq:1.8} 
  \meanx{(\nabla \ln P)^2 } = -\meanx{\nabla^2 \ln P}\;. 
\end{equation} 
As in Eq.~\eqref{eq:1.4} only the kinetic energy varies, one has $\delta \meanx{E_{{\rm kin}} }(t) = \delta \meanx{E_{{\rm kin}} }(0) ={\rm \; const}$. Thus, with the Gaussian \eqref{eq:1.3}, this provides an expression for the averaged fluctuating kinetic energy, or heat, of a particle and its surroundings,
\begin{equation} \label{eq:1.9} 
\begin{array}{l}
  \displaystyle
  \delta \meanx{E_{\rm kin} (t)} = \frac{m}{2} \meanx{(\delta v)^2 }+\frac{m}{2} \meanx{u^2 } 
      = \frac{m}{2} \meanx{(\delta v)^2 }+\frac{\hbar^2 }{8m\sigma^2} = 
  \\[1.5ex]
  \displaystyle
  = \delta \meanx{E_{\rm kin} (0)} = 0 + \frac{m}{2} \left.\meanx{u^2 }\right|_{t=0} 
      = \frac{\hbar^2 }{8m\sigma_0^2 } =: \frac{m}{2} u_0^2 \;. 
\end{array} 
\end{equation} 
Moreover, with the diffusion constant
\begin{equation} \label{eq:1.10} 
  D := \frac{\hbar}{2m}   
\end{equation} 
Eq.~\eqref{eq:1.9} provides an expression for the initial velocity fluctuation
\begin{equation} \label{eq:1.11} 
  u_0 = \frac{D}{\sigma_0 }\;. 
\end{equation}

However, in a next step we now take into account the small momentum fluctuations $m\delta u$, providing an altered convective velocity $v\to v+\delta v(t)$, and thus an additional displacement $\delta x= \left|\delta u\right|t= \left|\delta v\right|t$, i.e., as soon as $t>0$. 
Therefore, returning to the simple rule of Eq.~\eqref{eq:1.12}, one now must decompose $u(t)$ into its initial value $u_0 $ and a fluctuating contribution $\delta u(t)$, respectively. Unless some thermal equilibrium were reached, the latter is typically given off from the ``heated'' thermal bath to the particle of velocity $v$,
\begin{equation} \label{eq:1.16} 
  u(t) = u_0 -\delta u(t)\;. 
\end{equation} 
As is detailed in ref.~\cite{Groessing.2010emergence}, this leads ultimately to the substitution of $u$ by $u_0$ in Eq.~\eqref{eq:1.15}, i.e.,
\begin{equation} \label{eq:1.17} 
  \meanx{x^2(t)} = \meanx{x^2(0)} + u_0^2 t^2 \;. 
\end{equation} 
Inserting \eqref{eq:1.11} into \eqref{eq:1.17} for the particular case that $\meanx{x^2(t)} \equiv \sigma^2 $ (i.e., $\meanx{x^2(0)} \equiv \sigma_0^2 $) provides for the time evolution of the wave packet's variance
\begin{equation} \label{eq:1.18} 
  \sigma^2 = \sigma_0^2 \left(1+\frac{D^2 t^2 }{\sigma_0^4 } \right)\;. 
\end{equation} 
The quadratic time-dependence of the variance $\sigma^2 $ is remarkable insofar as in ordinary diffusion processes the scenario is different. There, with the Gaussian distribution being a solution of the heat equation, for purely Brownian motion the variance grows only linearly with time, i.e., as described by the familiar relation
\begin{equation} \label{eq:1.19} 
  \meanx{x^2(t)} = \meanx{x^2(0)} + 2Dt\;. 
\end{equation} 
However, as we have seen, the momentum exchange between the particle and its environment is characterized by both a changing velocity and by a changing thermal environment of the particle, i.e., also by a changing diffusivity. Therefore, Eq.~\eqref{eq:1.19} must be modified to allow for a time-dependent diffusivity. In other words, we shall have to deal with the field of \textit{anomalous diffusion}. (For a short introduction, see, for example, ref.\ \cite{vainstein.2008}.) This means that instead of the diffusion constant $D$ in Eq.~\eqref{eq:1.2}, we now introduce a time-dependent diffusion coefficient $D(t) =kt^{\alpha } $, where $k$ is a constant factor. The exponent $\alpha $ was derived \cite{Groessing.2010entropy} to be $\alpha =1$. Therefore, 
$k = D^2/\sigma_0^2 = u_0^2$, 
and the time-dependent diffusion coefficient becomes
\begin{equation} \label{eq:1.24} 
  D(t) = u_0^2 \, t = \frac{D^2 }{\sigma_0^2 } \, t = \frac{\hbar^2 }{4m^2 \sigma_0^2 }\, t \;. 
\end{equation} 
Note that with the exponent of $t$ being $\alpha =1$, or the $t^2 $-dependence of $\sigma^2 $ in \eqref{eq:1.18}, respectively, one deals with the special case of anomalous diffusion usually named \textit{ballistic diffusion}. We shall review some general properties of ballistic diffusion at the end of the paper. At this point, however, it is useful to recall that throughout our modelling of sub-quantum processes, we deal with various processes at different time scales. On the shortest scales, we have assumed Brownian-type motions (not detailed here), which, on the next higher level of (spatial and) temporal scales lead collectively to the emergence of a regular diffusion wave. The latter is characterized by a velocity $u$ according to \eqref{eq:1.12}, and it is orthogonal on average to the particle's velocity $v$, thus providing the r.m.s.\ displacement \eqref{eq:1.15} depending on $u(t)$. As a next step, we have introduced the noisy thermal bath of the particle's environment, i.e., essentially the effect of other diffusion wave configurations, which disturbs the relation \eqref{eq:1.15} by introducing a fluctuating term $\delta u$. The net effect of the latter, however, is the r.m.s.\ displacement \eqref{eq:1.17} with a dependence solely on the initial diffusive velocity $u_0 $. This manifests itself also in the expression for $D(t)$ of the ultimately emerging ballistic diffusion, which is also dependent only on $u_0 $. 
However, even on this level of ballistic diffusion one can recover the signature of Brownian motion. In fact, if one considers the time-average of $D(t)$ for large enough times 
$t \gg 1/\omega$, i.e.,
\begin{equation} \label{eq:1.25} 
  \meant{D(t)} := \frac{1}{t} \int_0^t D(t') \d t' = \frac{u_0^2 }{2} \, t = \frac{D(t)}{2} \;, 
\end{equation} 
one immediately obtains the linear-in-time Brownian relation
\begin{equation} \label{eq:1.26} 
  \meanx{x^2(t)} = \meanx{x^2(0)} + 2\meant{D(t)} \, t \quad\text{and}\quad \sigma^2 = \sigma_0^2 + 2\meant{D(t)} t\;, 
\end{equation} 
which is, however, also in accordance with the $t^2$--dependence of Eq.~\eqref{eq:1.17}.

Note that the diffusivity's rate of change is a constant,
\begin{equation} \label{eq:1.27} 
  \frac{\d D(t)}{\d t} = \frac{D^2 }{\sigma_0^2 } = u_0^2 = {\rm const.}, 
\end{equation} 
such that it is determined only by the initial r.m.s.\ distribution $\sigma_0 $, providing also a re-formulation of Eq.~\eqref{eq:1.17} as
\begin{equation} \label{eq:1.28} 
  \meanx{x^2(t)} = \meanx{x^2(0)} + \frac{\d D(t)}{\d t}\, t^2 \;. 
\end{equation} 
With the square root of \eqref{eq:1.18},
\begin{equation} \label{eq:1.29} 
  \sigma = \sigma_0 \, \sqrt{1 + \frac{D^2 t^2 }{\sigma_0^4 } }\;,
\end{equation} 
we note that $\sigma/\sigma_0$ is a spreading ratio for the wave packet independent of $x$. This functional relationship is thus not only valid for the particular point $x(t) = \sigma (t)$, but for all $x$ of the Gaussian. Therefore, one can generalize \eqref{eq:1.29} for all $x$, i.e.,
\begin{equation} \label{eq:1.30} 
  x(t) = x(0)\frac{\sigma}{\sigma_0}, \quad\text{where}\quad \frac{\sigma}{\sigma_0} = \sqrt{1 + \frac{D^2 t^2}{\sigma_0^4}} \;. 
\end{equation} 

Now we remind ourselves that we deal with a particle of velocity $v={p\mathord{\left/ {\vphantom {p m}} \right. \kern-\nulldelimiterspace} m} $ immersed in a wave-like thermal bath that permanently provides some momentum fluctuations $\delta p$. The latter are reflected in Eq.~\eqref{eq:1.29} via the r.m.s.\ deviation $\sigma (t)$ from the usual classical path. In other words, one has to do with a wave packet with an overall uniform motion given by $v$, where the position $x_0 =vt$ moves like a free classical particle. As the packet spreads according to Eq.~\eqref{eq:1.29}, $x(t) = \sigma (t)$ describes the motion of a point of this packet that was initially at $x(0) = \sigma_0 $. Depending on whether 
initially $x(0)>\sigma_0 $ or $x(0)<\sigma_0 $, then, respectively, said spreading happens faster or slower than that for $x(0) = \sigma_0 $. In our picture, this is easy to understand. For a particle exactly at the center $x_0 $ of the packet [$x(0) =0$], the momentum contributions from the ``heated up'' environment on average cancel each other for symmetry reasons. 

However, the further off a particle is from that center, the stronger this symmetry will be broken, i.e., leading to a position-dependent net acceleration or deceleration, respectively, or, in effect, to the ``decay of the wave packet''. Moreover, also the appearance of the time-dependent diffusivity $D(t)$ is straightforward in our model.
Essentially, the ``decay of the wave packet'' simply results from sub-quantum diffusion with a diffusivity varying in time due to the particle's changing thermal environment: as the heat initially concentrated in a narrow spatial domain gets gradually dispersed, so must the diffusivity of the medium change accordingly.

Finally, one obtains with Eqs.~\eqref{eq:1.30} and \eqref{eq:1.10} for the ``smoothed out'' \textit{trajectories} (i.e., those averaged over a very large number of Brownian motions) 
\begin{equation} \label{eq:1.31} 
  x_{\rm tot} (t) = v t + x(t) = v t + x(0)\frac{\sigma}{\sigma_0} = v t + x(0)\sqrt{1+\frac{\hbar^2 t^2 }{4m^2 \sigma_0^4}} \;. 
\end{equation} 
Also, one can now calculate the \textit{average total velocity of a Gaussian wave packet}, 
\begin{equation} \label{eq:1.32} 
  v_{\rm tot} (t) = \frac{\d x_{\rm tot} (t)}{\d t} = v(t) + \frac{\d x(t)}{\d t} \;, 
\end{equation} 
providing \cite{Groessing.2010emergence}
\begin{equation} \label{eq:1.33} 
  v_{\rm tot} (t) = v(t) + \left[x_{\rm tot} (t) - v t\right]\frac{\hbar^2}{4m^2}\, \frac{t}{\sigma^2\sigma_0^2 } \;. 
\end{equation}

\section{Simulation of quantum phenomena with purely classical diffusion using coupled map lattices}

It is straightforward to simulate the diffusion process of Eq.~\eqref{eq:1.26} in a simple computer model. Using coupled map lattices (CML), one approximates the heat equation as usual by
\begin{equation} \label{eq:1.34} 
  P[i,k+1] = P[i,k] + \frac{D[i,k]\Delta t}{\Delta x^2} \left\{P[i+1,k] - 2P[i,k] + P[i-1,k]\right\} \;, 
\end{equation}
with space and time grid indices $i$ and $k$, respectively. For our anomalous (``ballistic'') diffusion one simply inserts \eqref{eq:1.24} into \eqref{eq:1.34}.

\begin{figure}[!t]
	\begin{minipage}[t]{7.0cm}
		\includegraphics[width=74mm,height=117mm]{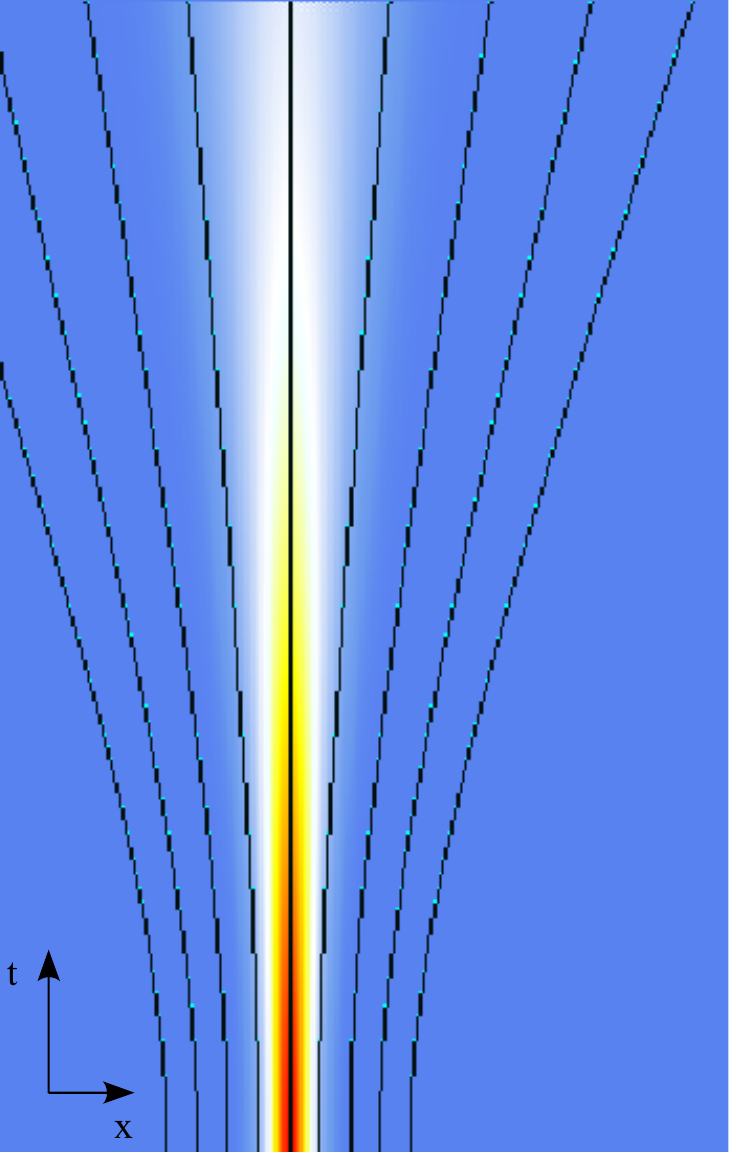}
		\caption{Classical CML simulation of the dispersion of a free Gaussian wave packet.}
		\label{fig:1}
	\end{minipage}
	\rule{13mm}{0pt}
	\begin{minipage}[t]{7.0cm}
		\includegraphics[width=74mm,height=117mm]{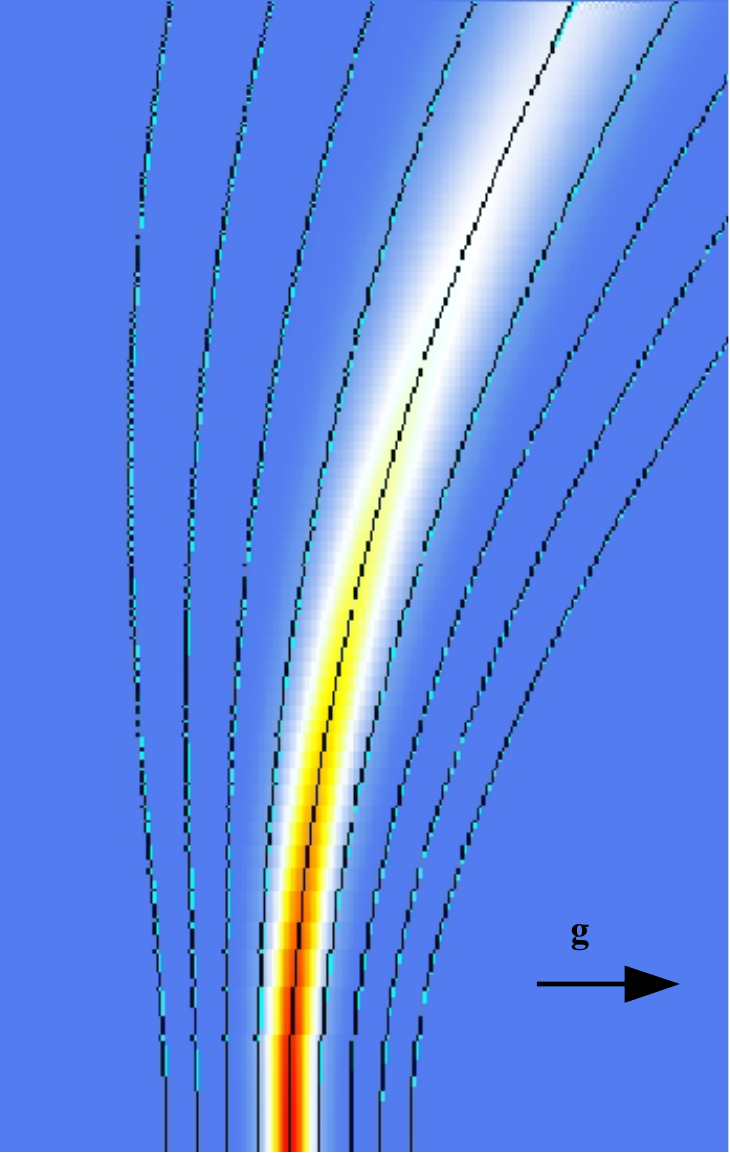}
		\caption{Dispersion of a Gaussian wave packet in a linear potential (e.g., a gravitational field).}
		\label{fig:2}
	\end{minipage}
\end{figure}

The result is depicted in Fig.~\ref{fig:1}, where the (macroscopic, classical) velocity is chosen as $v=0$. (For examples with $v\ne 0$ and different $\sigma_0 $, see \cite{Groessing.2010emergence}.) Moreover, nine exemplary averaged Bohmian trajectories are shown in Fig.~\ref{fig:1}, and it must be stressed that \textit{the Figure shows the emerging behaviour of the Gaussian packet following solely from the CML simulation of Eq.~\eqref{eq:1.34}}. In addition, the emerging trajectories from the simulation (dark lines) are shown together with the calculated ones from Eq.~\eqref{eq:1.31} (bright, mostly superposed by the dark lines), providing \textit{exactly the same trajectories} (i.e., up to resolution limits due to discretization).

Note that the trajectories are not the ``real'' ones, but only represent the averaged
behaviour of a statistical ensemble. The results are in full concordance with quantum theory, and in particular with Bohmian trajectories. (For a comparison with the latter, see, for example, \cite{Holland.1993}, or the Figures for the Gaussian wave packet example in \cite{von_bloh.2010}, which are in excellent agreement with our Fig.~\ref{fig:1}.) This is so despite the fact that no quantum mechanics has been used yet, i.e., neither a quantum mechanical wave function, or the Schr\"odinger equation, respectively, nor a guiding wave equation, nor a quantum potential. Moreover, we want to stress that our model offers possible insights into the sub-quantum domain which must escape (Bohmian or orthodox) quantum theory because the latter simply does not employ the ``language'' necessary to express them. Note, for example, that the existence of the hyperbolic trajectories depicted in Fig.~\ref{fig:1}, which are given by the formula for the scale invariant wave packet spread \eqref{eq:1.18}, has a simple physical explanation in terms of sub-quantum processes. As the inflection points of the hyperbolas are, according to \eqref{eq:1.18}, characterized by the relation 
$D^2 t^2 / \sigma_0^4 \cong 1$, 
i.e., by the length scales $u_0^2 t^2 \cong \sigma_0^2 $, the trajectories' evolution is easily understood: as long as the main bulk of the heat ``stored'' in the 
initial Gaussian spreads well ``inside'' the distribution, $u_0^2 t^2 <\sigma_0^2 $, the average particle velocity $v$ is not affected much. However, if said main bulk approximately reaches the distance $\sigma_0 $, or spreads to regions $u_0^2 t^2 >\sigma_0^2 $, respectively, the particles will ``feel'' the full heat and get propagated into new directions. For $t\to \infty $, then, $u_0 $ becomes the spreading rate of the whole Gaussian packet:
\begin{equation} \label{eq:1.35} 
  \frac{\d\sigma}{\d t} = \frac{\hbar^2 t}{4m^2 \sigma_0^2\sigma} \; \stackrel{t\to \infty }{\longrightarrow} \; 
      \frac{\hbar}{2m\sigma_0} = u_0 \;. 
\end{equation} 
In other words, the ``spreading'' already begins at $t=0$, but becomes ``visible'' in terms of deflected trajectories only when
$t\cong {\sigma_0 \mathord{\left/ {\vphantom {\sigma_0  u_0 }} \right. \kern-\nulldelimiterspace} u_0 } $.

So far we have shown that free one-particle quantum motion is identical to sub-quantum ballistic diffusion. This is the basis of a research program that would eventually cover more and more complex situations beyond the case of free motion. As a first simple example, we extend the present scheme to include a linear potential. That is, we place the initial Gaussian packet \eqref{eq:1.3} in a uniform potential $V=K\cdot x$, which may be an electric or a gravitational field, for example. For illustration, but without loss of generality, we substitute in the following $K$ by $mg$, i.e., we deal with a Gaussian packet freely falling due to the potential $V=mg\cdot x$.

After a short calculation \cite{Groessing.2010entropy}, one obtains
the trajectories of particles in a gravitational field,
\begin{equation} \label{eq:2.3} 
  x_{\rm tot} (t) = v t - \frac{g}{2} \, t^2 + x(0)\sqrt{1+\frac{\hbar^2 t^2}{4m^2 \sigma_0^4}} \;,
\end{equation} 
and the particle acceleration
\begin{equation} \label{eq:2.4} 
  \ddot{x}_{\rm tot} = -g + x(0)\frac{u_0^2}{\sigma_0^2} \left[1+\frac{u_0^2 t^2}{\sigma_0^2} \right]^{-3/2} \;.
\end{equation}

In Fig.~\ref{fig:2}, exemplary trajectories of a Gaussian in a gravitational field are shown as obtained by the CML simulation of ballistic diffusion, modified by the substitution  $v\to v-gt$. The trajectories (dark lines) superpose those (bright lines) derived directly from 
Eq.~\eqref{eq:2.3}, and again exhibit excellent agreement.

Note that some trajectories of the dispersing Gaussian even overcome ``gravity'' for a well-defined period of time, as can also be seen in Fig.~\ref{fig:2}. In fact, our sub-quantum model provides a detailed explanation of why, and within which time limits, this ``anti-gravity'' effect becomes possible. A look at the last term of Eq.~\eqref{eq:2.4} provides the answer. Similarly to the discussion of the hyperbolas' inflection points in the free case, one deals also here with an extremum at the scale $u_0^2 t^2 \cong \sigma_0^2 $. However, this time the corresponding expression (in rectangular brackets) is antagonistic to $g$. In other words, said scale describes the maximum of the ``anti-gravity'' effect, because it is there where the heat of the main bulk of the packet is consumed, which has via the kinetic energy counter-acted the effect of gravity. For larger times, then, the remaining heat gets gradually less, and therefore gravitational acceleration begins to dominate the trajectories' curvature.

\begin{figure}[!t]
	\begin{minipage}[t]{7.0cm}
		\includegraphics[width=70mm,height=117mm]{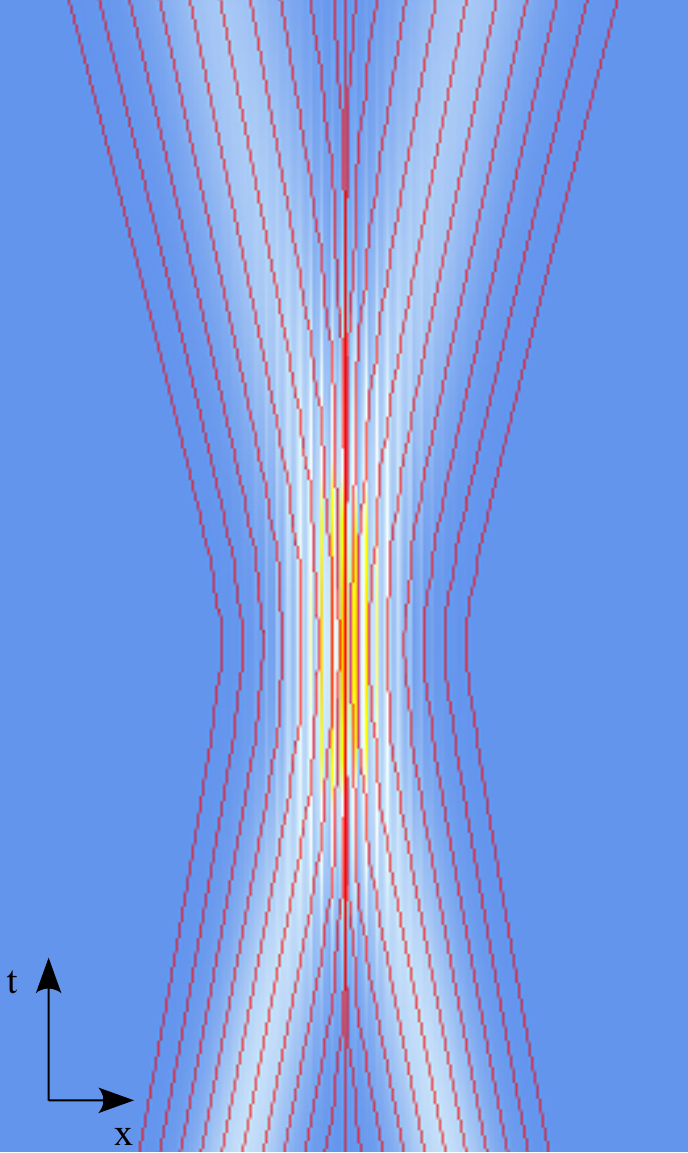}
		\caption{Trajectories of two Gaussians with the same half widths.}
		\label{fig:3}
	\end{minipage}
	\rule{18mm}{0pt}
	\begin{minipage}[t]{7.0cm}
		\includegraphics[width=69mm,height=117mm]{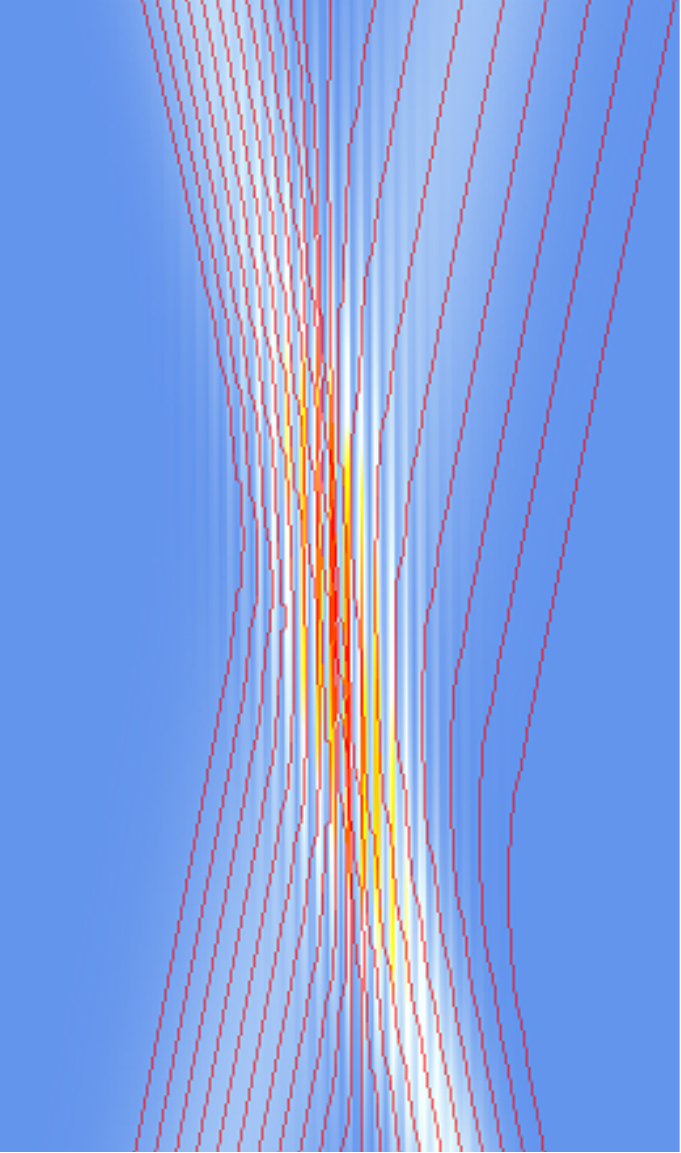}
		\caption{Trajectories of two Gaussians with half width ratio of 2:1.}
		\label{fig:4}
	\end{minipage}
\end{figure}

Finally, instead of following just one Gaussian, we extend our simulation scheme to include two possible paths of a particle which eventually cross each other. For this, we use two Gaussians approaching each other. In the space-time diagram of Fig. 3, they are characterized by the same half-widths $\sigma$, whereas in Fig.~\ref{fig:4} we choose Gaussians whose half-widths' ratio is 2:1. 

Note that in both Figs.~\ref{fig:3} and \ref{fig:4} one can observe a basic characteristic of the (averaged) particle trajectories, which, just because of the averaging, are identical with the Bohmian trajectories. To fully appreciate this characteristic, we remind the reader of the severe criticism of Bohmian trajectories as put forward by Scully and others 
(see \cite{Scully.1998}, and references therein.) The critics claimed that Bohmian trajectories would have to be described as ``surreal'' ones because of their apparent violation of momentum conservation. In fact, due to the ``no crossing rule'' for Bohmian trajectories in Young's double slit experiment, for example, the particles coming from, say, the right slit (and expected at the left part of the screen if ``classical'' momentum conservation should hold) actually arrive at the right part of the screen (and vice versa for the other slit). In Bohmian theory, this ``no crossing rule'' is due to the action of the non-classical quantum potential, such that, once the existence of a quantum potential is accepted, no contradiction arises and the trajectories may be considered ``real'' instead of ``surreal''. 

Here we can note that in our sub-quantum approach an explanation of the ``no crossing rule'' is even more straightforward and actually a consequence of a detailed microscopic momentum conservation. As can be seen particularly clearly in Fig.~\ref{fig:3}, the (Bohmian) trajectories are repelled from the central symmetry line. However, in our case this is only implicitly due to a ``quantum potential'', but actually due to the identification of the latter with a kinetic (rather than a potential) energy: it is the ``heat of the compressed vacuum'' that accumulates along said symmetry line (i.e., as reservoir of ``outward'' oriented  kinetic energy) and therefore repels the trajectories. In Fig.4, moreover, one sees the resulting trajectories for the case that the symmetry of Fig.3 is broken: by using different half-widths for the two paths, one obtains the skew-symmetric distribution of particle intensities and trajectories, respectively. Both Figs.~\ref{fig:3} and \ref{fig:4} are in full concordance with Bohmian trajectories (cf., for example, \cite{Sanz.2008} for comparison). However, as mentioned, in our case also a micro-causal explanation is provided, which brings the whole process into perfect agreement with momentum conservation on a more ``microscopic'' level.

To sum up, we have shown for the cases of free motion and motion in linear potentials, respectively, that the time evolution of a one-particle quantum system in the noisy heat bath of the surrounding ``vacuum'' exactly equals that of (classical) ballistic sub-quantum diffusion.  Note that there are some well-known general characteristics of ballistic diffusion \cite{vlahos.2008}, and the results presented in this paper agree perfectly with them. For one thing, ballistic diffusion is the only type of diffusion that exhibits reversibility, and because of this it violates ergodicity (i.e., as in our cases). Also, if the ballistic system is not in equilibrium initially, it will never reach equilibrium (which is true here as well). Finally, the result of any measurement depends on the initial conditions. This can be clearly seen also from our results for the time evolution of the Gaussians and the corresponding averaged trajectories, which all depend on the initial values of $u_0$, or $\sigma_0$, respectively.

\section*{References}

\providecommand{\newblock}{}

\end{document}